\documentstyle[prd,preprint,aps]{revtex}

\newcommand{\be}{\begin{equation}}
\newcommand{\ee}{\end{equation}}
\newcommand{\bea}{\begin{eqnarray}}
\newcommand{\eea}{\end{eqnarray}}

\newcommand{\A} {{\cal A}}
\newcommand{\E} {E_0}
\newcommand{\EP} {E_0^{\prime}}
\newcommand{\EPP} {E_0^{\prime\prime}}

\newcommand{\LL} {{\cal L}}
\newcommand{\ZG} {{\cal{Z}}_G}
\newcommand{\GG} {G_0}
\newcommand{\GP} {G_0^{\prime}}
\newcommand{\GPP} {G_0^{\prime\prime}}
\newcommand{\GD} {{\dot G}_0}
\newcommand{\GDD} {{\ddot G}_0}
\newcommand{\ED} {{\dot E}_0}
\newcommand{\EDD} {{\ddot E}_0}
\newcommand{\EDP} {{\dot E}_0^{\prime}}
\newcommand{\GDP} {{\dot G}_0^{\prime}}

\newcommand{\XXXX} {\left(an +b m^p\right)}

\newcommand{\YYYY} {\left(an_0 +b m_0^p\right)}
\newcommand{\BBB} {\left(n+\alpha m^p\right)}
\newcommand{\QQ} {\Psi_{0}}

\begin{document}

\reversemarginpar
\tighten

\title{Thermal Fluctuations and  Black Hole Entropy}

\author{Gilad Gour\thanks{E-mail:~gilgour@phys.ualberta.ca}
and A.J.M. Medved\thanks{E-mail:~amedved@phys.ualberta.ca}}

\address{
Department of Physics and Theoretical Physics Institute\\
University of Alberta\\
Edmonton, Canada T6G-2J1\\}

\maketitle

\begin{abstract}

In this paper, we consider the effect of thermal fluctuations 
on the entropy of both  neutral and charged black holes. 
We emphasize the distinction between  fixed and fluctuating
charge systems;  using a canonical ensemble to describe the former
and  a grand canonical ensemble to study the latter. 
Our novel approach is based on the philosophy
that the black hole quantum spectrum is an essential
component in any such calculation. For definiteness, we
employ a uniformly spaced area spectrum, which has been
advocated by Bekenstein and others in the literature.
 The generic results are applied to
 some specific models; in particular,
various limiting cases of an (arbitrary-dimensional)
  AdS-Reissner-Nordstrom
black hole.  We find that
the leading-order quantum correction to the entropy
can consistently  be  expressed as the logarithm of the 
classical quantity.
For a  small AdS curvature parameter and  zero net charge,
it is shown that, independent of the dimension, 
the logarithmic prefactor is +1/2  when the charge is
fixed but +1 when the charge is fluctuating.
We also demonstrate that, 
in the grand canonical framework, the fluctuations in the 
charge are large, $\Delta Q\sim\Delta A\sim S_{BH}^{1/2}$,
even when $\langle Q\rangle =0$.  A further implication
of this framework is that
an asymptotically flat,  non-extremal  black hole
can never achieve  a state of thermal equilibrium.

\end{abstract}

\section{Introduction}

A popular notion in modern research is that some fundamental
theory, commonly  referred to as {\it quantum gravity},
will be necessary to describe physics at energy scales
in excess of the Planck mass.  Unfortunately, 
any progress in quantum gravity is severely constrained
by a simple fact: the relevant scales can
not be probed experimentally (at least not directly).
 It, therefore, becomes important to  search for
 criteria that can be used to test the
viability of a prospective fundamental theory 
\cite{smo}.

One such viability test is an explanation of  the
Bekenstein-Hawking black hole
 entropy \cite{bek1,haw1}. To elaborate, thermodynamic
arguments imply that a black hole has an entropy of 
\be 
S_{BH}={A\over 4 G}
\label{111}
\ee
 (where  $A$ is the horizon surface area  and
 $G$ is Newton's constant\footnote{Here and throughout, 
all  fundamental constants, besides $G$, are set to unity.}); 
however, the statistical origin for this entropy
remains conspicuously unclear. Presumably, one
can calculate this entropy, in principle, by
tracing over an appropriate set of  fundamental degrees of freedom. 
That is to say, any acceptable theory of quantum gravity
should  be able to reproduce, at the level of microstate counting,
the quantitative relation between $S_{BH}$ and $A$.

There has, unquestionably, been substantial success along the above lines
\cite{wald1}.
For instance, both string theory \cite{str1} and loop quantum gravity
\cite{ash}
have (at least under certain conditions)  reproduced
the black hole area law~(\ref{111}).
In fact, this remarkable agreement between
significantly different approaches would
suggest that further discrimination is required. In this regard, 
it has been proposed \cite{maj1x} that one should  examine the
 quantum corrections
to the {\it classical} area law (\ref{111}).
(Note that, regardless of its fundamental
origins, $S_{BH}$ arises, thermodynamically, at the
tree level.) 

There has, indeed, been much recent interest in
calculating the quantum corrections to $S_{BH}$.
Various approaches have been utilized for this purpose; including 
methodologies based on, for example,
Hamiltonian partition functions \cite{makx},
 loop quantum gravity \cite{maj1x},
near-horizon symmetries \cite{carx} and general thermodynamic
arguments \cite{das2x}. (Also see
\cite{sol1x,furx,loux,frox,sol2x,kasx,%
kun1x,gg1x,ostx,jinx,das1x,bir1x,maj2x,kun2x,ms2x,bir2x,govx,%
cavx,gup1x,ajm1x,gup2x,mukx,das3x,ajm2x,gg2x,kripx,%
kaul1x,kaul2x,maj3x,das4x,das5x}.) One common characteristic 
of {\it all} the cited methods is a leading-order
correction that is proportional to $\ln{S_{BH}}$. Nonetheless,
the proportionality constant - that is, the value of  the logarithmic
prefactor - does not exhibit the same  universality.

 This apparent discrepancy in   the prefactor
can be partially explained by a point that is not always
clarified in the literature: there are, in fact, 
 {\it two} distinct and separable
 sources for this logarithmic correction
 \cite{gg2x,maj3x}.
Firstly, there should be  a  correction to the number
of microstates that are necessary to describe a black hole
of fixed horizon area. That is, a quantum-correction 
 to the {\it microcanonical}
entropy,\footnote{The horizon area of a (for instance) 
four-dimensional black hole can 
be expressed as $A=16\pi G^2 E^2$, where $E$ is the black hole
conserved energy. Hence, a fixed area translates into a fixed energy,
and a  microcanonical framework is, therefore, the appropriate one.} 
which  can, on general heuristic grounds, 
be expected to be negative.
Secondly, as any black hole will typically exchange heat (or matter) with its
surroundings,
 there should also be a  correction due to thermal fluctuations
in the horizon area. That is, a  {\it canonical} correction 
that must certainly be positive, as it increases the uncertainty
of the horizon area (and, thus, the entropy).

One might anticipate that only the former, microcanonical
correction should depend on the fundamental 
degrees of freedom;
that is,  the thermal correction should be obtainable 
from  some canonical analysis that makes
no direct reference to  quantum gravity. In a limited sense,
this may still  be true; however, we will
argue below that a certain  aspect of quantum gravity
- namely, the black hole quantum spectrum -
 does
indeed enter into the thermal calculation. That this detail  
has been  neglected, in some of  the recent literature,
is a central motivation for the current work.

To help illustrate our point, we will briefly review a 
calculation of the thermal correction, as presented
in a recent canonical treatment by Chatterjee and
Majumdar \cite{maj3x}.\footnote{For closely related
works, also see \cite{das2x,das3x,das4x}. Note, as well, that
our notation differs somewhat from that of \cite{maj3x}.}
These authors essentially start with a
standard canonical partition function,
\be
{\cal Z}_C(\beta)=\int^{\infty}_{0}dE g(E)e^{-\beta E}\;,
\label{112}
\ee
where $\beta^{-1}$ is the fixed temperature, $E$ is
the energy and $g(E)$ is the density of states. 
They also make the usual identification,
\be
g(E)=e^{S(E)}\;,
\label{113}
\ee
where $S(E)$ is the microcanonical entropy. 
Expanding  $S(E)$ about the equilibrium value
of energy ($E_0$) and imposing $\beta=\partial_{E}S(E_0)$
(via the first law of thermodynamics), the authors
obtain a Gaussian integral. This yields
\be
{\cal Z}_C (\beta) \approx e^{-\beta E_0 +S(E_0)}
\left[2\pi\over -\partial_E^2S(E_0)\right]^{1\over 2}\;.
\label{114}
\ee
Using textbook statistical mechanics,
they  find the following expression
for the canonical entropy:
\be
S_C \approx S(E_0) -{1\over 2}\ln\left[-\partial_E^2S(E_0)\right]\;.
\label{115}
\ee
It is then possible to identify $S(E_0)$ as the black
hole entropy  $S_{BH}$ (up to the previously discussed
microcanonical correction, which is inconsequential
to the current discussion)  and the logarithmic term
as the leading-order correction due to
thermal fluctuations.  
It is straightforward and  useful to apply 
this formalism  to   an explicit example;
for instance, the BTZ black hole \cite{btz}.\footnote{A detailed
discussion of this model can be found in  Section III of the current paper.}
In this case \cite{maj3x},
\be
S_C\approx S_{BH}+{3\over 2}\ln S_{BH}\;.
\label{116}
\ee

Let us now  address the issue at hand.  The canonical
partition function (\ref{112}) should really
be viewed as the continuum limit of a discrete
sum. We can, quite generically, express the partition function as
\be
{\cal Z}_C(\beta)=\sum_{i}  g_i  e^{-\beta E_i}\;,
\label{117}
\ee
where $i$ is whatever quantum numbers label the energy levels of the
black hole and $g_i$ is the degeneracy of a given level.
However, it should be clear that Eq.(\ref{112}) can only follow
from Eq.(\ref{117}) if the energy levels are evenly spaced
(and, if anything, one would expect an evenly spaced area
spectrum - see below). More generally, the continuum
limit of Eq.(\ref{117}) would lead to
\be
{\cal Z}_C(\beta)=\int^{\infty}_{0}dE 
\left[{\partial_i E}\right]^{-1} 
g(E) e^{-\beta E}
\label{118}
\ee
and, consequently, a canonical entropy of
\be
S_C \approx S(E_0) -{1\over 2}\ln\left[-\partial_E^2S(E_0)\right] 
-\ln\left[\partial_i E(E_0)\right]\;.
\label{500}
\ee
It is now evident that this extra factor of
$\left[\partial_{i}E\right]^{-1}$ - 
or the ``Jacobian'' -   enters the canonical entropy at 
precisely the logarithmic
order.  

To emphasize our point, let us reconsider the BTZ black
hole and, for the sake of argument, assume an  evenly spaced area
spectrum (that is, $i\rightarrow A$).  A simple calculation
reveals that $\partial_A E\sim A\sim S_{BH}$, and so
Eq.(\ref{116}) should be modified as follows: 
\be
S_C\approx S_{BH}+{1\over 2}\ln S_{BH}\;.
\label{501}
\ee
Of course, Eq.(\ref{116}) could still be the valid result,
depending on the true nature of black hole spectroscopy. 
Our main point is that such spectral considerations
must be dealt with and
can {\it not} be disregarded {\it a priori}. 

The primary focus of the current paper is
to calculate the thermally induced corrections
to the (classical) black hole entropy. As discussed above, such
calculations may ultimately have relevance as a means of
discriminating  candidates for the fundamental theory.
(In this regard, the microcanonical corrections may be
of even greater interest;  however, except for a few comments,
this part of the calculation will not be addressed here.)
Unlike prior works along this line, we will 
directly be incorporating the effects of black hole
spectroscopy. For definiteness,  
 a uniformly spaced area spectrum will be employed throughout. 
Although somewhat conjectural, this form of spectrum
has been strongly advocated in the literature;
beginning with the heuristic arguments of
Bekenstein \cite{bek2,muk,bek3}. More recently (and more rigorously),
this spectrum has received support from  Bekenstein's
algebraic approach to black hole quantization \cite{bek5,bek6,gg2x},
the reduced phase space approach initiated by Barvinsky and Kunstatter
\cite{kun1,kun2,kun3,das1,gj1,gj2}, and the WKB treatment
of Makela and others \cite{mak1}.\footnote{Yet more
references  can be found in \cite{bek6}. Furthermore, see 
\cite{lgq} for favorable arguments in  the context of loop quantum gravity.}
Furthermore, we suggest that the elegance of our results
may be viewed as further, independent support for
the evenly spaced area spectrum.

A further novelty of the current analysis is
that  an important distinction will be made between
black holes with a fixed  (electrostatic)
charge and those with a fluctuating charge. 
The latter  case of a dynamical charge  necessitates that the system be
modeled as a {\it grand canonical} ensemble. 
Although this scenario poses many  new technical challenges,
it  provides a much sterner test for   
the viability of the proposed spectrum and (as elaborated on
in the final section) is a necessary
step  towards a realistic treatment of the problem.

The remainder of the paper is organized as follows.
In the next section, we develop the general canonical
formalism as appropriate  for  black holes  with a fixed charge.
In Section III, we apply these generic results to 
some definite models; in particular,  various
limiting cases of an arbitrary-dimensional, anti-de Sitter (AdS), 
stationary black hole.
In Section IV, we regard the charge  as a fluctuating quantity and
 accordingly readdress the problem in a grand canonical
framework.  Special models are again used to
illustrate the (revised) formalism in Section V.
Finally, Section VI contains a summary and some
further discussion.

\section{Canonical Ensemble: General}

Our premise will be  a  black hole in a  ``box'';  that is, 
a black hole which is (up to fluctuations) in a state of thermal
equilibrium with its surroundings. The system can, therefore,
be modeled as  a canonical ensemble of particles and fields.
An appropriate form of  canonical 
partition function can be written as
\be
{\cal Z}_C(\beta)=\sum_{n} g_n \exp\left(-\beta E(n)\right)\;,
\label{119}
\ee
where $\beta^{-1}$ is the (fixed) temperature of the heat bath and
$n$ is a quantum number (or numbers) that parameterizes the black hole
spacetime. Also, $E(n)$ and $g_n$ represent the
energy and degeneracy of the $n$-th level.\footnote{Technically
speaking, the partition function, ${\cal Z}_C$, should also
be a function of the box size. When we consider
specific AdS models in the subsequent sections, the box size
will effectively  enter the calculations in the guise
of the AdS curvature parameter, $L$.}

As advertised  in the introductory section,  we
will adopt the well-motivated choice of
an evenly spaced area spectrum: $A(n)\sim n$ ($n=0,1,2,...$).
Equivalently, by  virtue of
the black hole  area law \cite{bek1,haw1}, 
$S(n)\sim n$.  Common-sense arguments  dictate that
$g_n \propto e^{S(n)}$, and so we can write
\be
\ln g_n =\epsilon n\;,
\label{120}
\ee
where $\epsilon$ is a positive, dimensionless parameter of
the order unity.

The partition function can now be expressed as
\be
{\cal Z}_C(\beta)
=\int^{\infty}_{0} dn \exp\left(-\beta E(n)+\epsilon n\right)\;,
\label{121}
\ee
where we have also taken the continuum limit. Such a limit
is
appropriate for a semi-classical ({\it i.e.}, large black hole) 
regime, which will always be our interest.

Ultimately, we also require an
explicit expression for
 the energy as a function of the spectral number. 
This can be achieved, for any
given black hole model, by way of  the first
law of black hole mechanics.
For the moment, let us keep matters
as general as possible and 
simply expand  the energy function about
$n_0 \equiv <n>$ (where $<...>$ denotes
the ensemble average):
\be
E(n)=E(n_0)+\left(n-n_0\right)E^{\prime}(n_0) +{1\over 2}
\left(n-n_0\right)^2 E^{\prime\prime}(n_0) +... \;,
\label{122}
\ee
with  a prime indicating  a derivative with respect
to $n$ (here and throughout).

Substituting the above expansion into the exponent of
Eq.(\ref{121}) and employing a trivial change
of integration variables, we obtain
\be
{\cal Z}_C(\beta)\approx \exp\left(-\beta \E +n_0\epsilon\right)
\int^{\infty}_{-n_0}dx \exp\left(-\beta\left[x\left(\EP
-{\epsilon\over\beta}\right)+{1\over 2}x^2 \EPP 
\right]\right)\;,
\label{123}
\ee
where $\E\equiv E(n_0)$, $\EP\equiv E^{\prime}(n_0)$,
{\it etcetera}.

In the semi-classical or large $n_0$ regime,
 the lower  limit can  be asymptotically extended ($-n_0\rightarrow -\infty$),
as any omitted terms (in the entropy) will
be of the order ${\cal O}[n^{-1}_{0}]$.  With this approximation
and another  shift in the integration variable,
we have a  Gaussian form,
\be
{\cal Z}_C(\beta)
\approx \exp\left[-\beta\E+n_0\epsilon+{\beta\over 2}{{\left(
\EP-{\epsilon\over\beta}\right)^2\over \EPP}}\right]
\int^{\infty}_{-\infty} dz \exp\left(-{1\over2}\beta \EPP z^2\right)\;,
\label{124}
\ee
which can be readily evaluated to yield
\be 
{\cal Z}_C(\beta)
\approx \exp\left[-\beta\E+n_0\epsilon+{\beta\over 2}{{\left(
\EP-{\epsilon\over\beta}\right)^2\over \EPP}}\right]
\sqrt{{2\pi\over \beta \EPP}}\;.
\label{125}
\ee
That is,
\be
\ln{\cal Z}_C\approx 
-\beta\E+n_0\epsilon + \left[{\beta\over 2}{{\left(
\EP-{\epsilon\over\beta}\right)^2\over \EPP}}\right]
-{1\over 2}\ln\left[\beta \EPP\right]\;.
\label{126}
\ee

We can now apply a textbook  thermodynamic relation, 
\be
S_C=\left(1-\beta{\partial\over\partial\beta}\right)\ln {\cal Z}_C\;,
\label{128}
\ee
to evaluate the canonical entropy:
\be
S_C\approx \epsilon n_0 -{\epsilon\over\EPP}\left(\EP-{\epsilon\over
\beta}\right)
-{1\over 2}\ln\left[\beta \EPP\right]\;.
\label{129}
\ee

By exploiting the first law of thermodynamics,
we will be able to  simplify the above outcome.  
First of all, in view  of Eq.(\ref{121}),
 $F(n)=\beta E(n) -\epsilon n$ can be identified  as
the microcanonical free energy. It follows that
$F^{\prime}(n_0)=0$, which translates into
\be
\EP ={\epsilon\over\beta}\;.
\label{1}
\ee
Hence, Eq.(\ref{129}) simplifies as follows:
\be
S_C\approx S_{BH}
-{1\over 2}\ln\left[\beta \EPP\right]\;.
\label{2}
\ee
Here, we have identified  $\epsilon n_0$
as the equilibrium value of the black hole
entropy, $S_{BH}$. (More generally, $S(n)=\epsilon n$.) 
The remaining term
represents the anticipated logarithmic correction.

A couple of comments are in order. Firstly,
it should be clear that the general procedure can only make sense
if the argument of the logarithm is strictly positive.
Since $\beta >0$ is universal  (assuming cosmic
censorship \cite{wald2}), this means that  $\EPP>0$
is a necessary (but not necessarily sufficient) constraint for 
attaining a state of thermal equilibrium.\footnote{A simple
calculation verifies that $\EPP >0$ is
equivalent to a positive specific heat.}
 This {\it stability}
condition (and its grand canonical analogue) will play a 
significant role in the subsequent analysis. 
Secondly, in the full quantum treatment, it would
be necessary to replace $S_{BH}$ with the microcanonical
entropy, which already contains a   quantum correction
of the  logarithmic order (see the prior section). 
That is, one would anticipate a canonical entropy
of the form $S_C=S_{BH} + \Delta_{MC} +\Delta_{C}$,
where $\Delta_{MC}$ represents the implied 
microcanonical correction and $\Delta_{C}$
represents the explicit (thermal) correction in Eq.(\ref{2}).
As  the current paper focuses on the consequences of
thermal fluctuations, we will continue to disregard $\Delta_{MC}$
until some comments in the final section. 
 
Our formal expression for $\ln Z_C$ can also be
used to quantify the thermal fluctuations in
the eigenvalue $n$ or, equivalently, the
variation in the black hole area. First
of all, let us confirm that $n_0$ is truly
the thermal expectation value of $n$.
This can be accomplished by standard techniques:
\be
<n>={1\over {\cal Z}_C}{\partial{\cal Z_C}\over \partial \epsilon}
={\partial\ln {\cal Z}_C\over\partial\epsilon}\;. 
\label{3}
\ee
Substituting  Eq.(\ref{126})  and employing
the equilibrium condition (\ref{1}), we do indeed  obtain
the anticipated result of $<n>=n_0$.

Next, let us evaluate $<n^2>$ by way of the following relation: 
\be
<n^2>={1\over {\cal Z}_C}{\partial^2 {\cal Z}_C\over \partial \epsilon^2}
={\partial^2\ln{\cal Z}_C\over \partial \epsilon^2}
+\left[{\partial\ln{\cal Z}_C\over \partial \epsilon}\right]^2\;.
\label{4}
\ee
Defining the variation ($\Delta n$) in the usual
way, we  have
\be
(\Delta n)^2 \equiv <n^2>-<n>^2=
{\partial^2\ln{\cal Z}_C\over \partial \epsilon^2}\;.
\label{5}
\ee
Some straightforward calculation then yields 
\be
(\Delta n)^2={1\over \EPP\beta}={\EP\over\epsilon\EPP}\;.
\label{6}
\ee

We can make sense of the last equation by  noting
that, typically (for large $n$), one can write $E\sim n^{\gamma}$, where
$\gamma$ is a model-dependent parameter. It follows that,
for a large class of black holes, 
\be
\Delta n\sim n_0^{1\over 2}\sim  S_{BH}^{1\over 2}\;,
\label{7}
\ee
as would be expected on intuitive grounds.

Finally, let us note that, by way of the area spectrum and  Eq.(\ref{6}),
the canonical entropy (\ref{2}) can  elegantly be  expressed
in terms  of  $\Delta S_{BH}$
({\it i.e.}, the fluctuations in the entropy or area):
\be
S_{C}\approx S_{BH}+\ln[\Delta S_{BH}]\;.
\label{entflu}
\ee

\section{Canonical Ensemble: Examples}

It is an instructive exercise to illustrate our
generic formalism with
 some specific black hole models. We will, in turn,
consider the BTZ black hole, AdS-Schwarzschild black holes
and AdS-Reissner-Nordstrom black holes. The latter
two cases will be carried out for a
spacetime of arbitrary dimensionality (more precisely, $d\geq4$).
For  all models under consideration, the AdS curvature
parameter, $L$,  can  be viewed as a measure of
the {\it effective} box size. Hence, the limit $L\rightarrow\infty$
can equivalently be regarded as either the limit of an 
asymptotically flat spacetime
or  an infinitely sized box.

\subsection{BTZ Black Hole}

The BTZ black hole \cite{btz} is a special solution of three-dimensional
AdS space that exhibits all of the usual properties 
 of a black hole spacetime. Besides being a useful ``toy'' model,
the BTZ black hole has sparked recent interest in the context
of  the AdS-CFT (conformal field theory) 
correspondence \cite{ads}.

By expressing the BTZ solution in a Schwarzschild-like gauge,
one can readily obtain the following relation between
the horizon radius, $R$, and the conserved energy, $E$ \cite{btz}:
\be
f(R)\equiv {R^2\over L^2}-8G_3E=0\;,
\label{130}
\ee
where $G_3$ is the three-dimensional Newton constant
and $L$ is the AdS$_3$ curvature parameter.
Hence,
\be
R=L\sqrt{8G_3 E}\;.
\label{131}
\ee

Let us next consider the three-dimensional  analogue to the
black hole area  law,
\be
S_{BH} ={A\over 4G_3}={2\pi R\over 4G_3} =\pi L\sqrt{2E\over G_3}\;.
\label{132}
\ee
By calling upon the spectral form of the entropy, $S(n)=\epsilon n$,
we can then express the energy as an explicit function
of $n$:
\be
E(n)= {G_3\over 2}\left[{\epsilon\over \pi L }\right]^2 n^2\;.
\label{134}
\ee

It follows that $\beta^{-1}\sim \EP\sim n_0$, $\EPP\sim{\rm constant}$,
and so $\beta\EPP\sim n_0^{-1}\sim S_{BH}^{-1}$. Substituting
into Eq.(\ref{2}), we find that
\be
S_C\approx S_{BH}+{1\over 2}\ln[S_{BH}]\;.
\label{138}
\ee
The logarithmic prefactor of +1/2 disagrees with the value
of +3/2  found
by (for instance) Chatterjee and Majumdar \cite{maj3x}.
 Nevertheless,
this discrepancy can be perfectly accounted for
by  incorporating the appropriate ``Jacobian'' (see the introductory
section) into their calculation.

\subsection{AdS-Schwarzschild}

We begin here with the defining relation for   the horizon radius
of a $d$-dimensional AdS-Schwarzschild black hole \cite{mys},

\be
f(R)\equiv {R^2\over L^2} +1-{\omega_d E\over R^{d-3}}=0\;,
\label{139}
\ee
or, solving for the energy, 
\be
E(R)={1\over \omega_d}\left[{R^{d-1}\over L^2}+R^{d-3}\right]\;,
\label{8}
\ee
where 
\be
\omega_d\equiv {16\pi G_d\over (d-2){\cal V}_{d-2}}\;.
\label{140}
\ee
In the above, $L$ now represents the $AdS_{d}$
curvature parameter,
 $G_d$ is the $d$-dimensional Newton constant, and
${\cal V}_{d-2}$ denotes the  volume of 
a $(d-2)$-dimensional spherical hypersurface (of unit
radius). 

As before, let us consider the black hole area law,
\be
S_{BH}={A\over 4G_d}={{\cal V}_{d-2} R^{d-2}\over 4G_d}\;,
\label{141}
\ee
and compare this to the entropic spectral form, $S(n)=\epsilon n$.
It directly follows that
\be
R^{d-2}(n)= \LL^{d-2} n\;,
\label{142}
\ee
where we have defined a convenient length parameter,
\be
\LL^{d-2}\equiv {(d-2)\epsilon\omega_d \over 4\pi}\;.
\label{143}
\ee

Substituting Eq.(\ref{142}) into Eq.(\ref{8}),
we obtain an explicit  spectral form for  the energy:
\be
E(n)= {1\over \omega_d}\left[{\LL^{d-1}\over L^2}n^{{d-1\over d-2}}
+\LL^{d-3} n^{{d-3\over d-2}}\right]\;.
\label{144}
\ee
The presence of two terms in the energy spectrum makes
a general calculation awkward; inasmuch as 
 the thermal correction to the entropy should, ideally, be
expressed 
as some prefactor times the logarithm of $S_{BH}$.
However, in certain limiting cases, such an
expression can readily  be attained, and we will proceed to
focus on a  pair of these special limits. 

{\bf (i) $L<<R$}:
For this limit of small box size, we can neglect 
the second term in Eq.(\ref{144}) and promptly obtain
$\E\sim n_0^{d-1\over d-2}$, 
$\beta^{-1}\sim \EP\sim n_0^{1\over d-2}$, and
 $\EPP=n_0^{-{d-3\over d-2}}$. Hence, 
$\beta \EPP \sim n_0^{-1}$ and Eq.(\ref{2}) reduces to
\be
S_C\approx S_{BH} +{1\over 2}\ln[S_{BH}]\;.
\label{150}
\ee

The logarithmic prefactor of +1/2 is again in disagreement
 with the value
of +1  found
in~\cite{maj3x} for a four-dimensional
AdS-Schwarzschild black hole (in the same limit). Moreover,
 the calculation in~\cite{maj3x} leads to a prefactor of +$d/2(d-2)$ when
 $d$ is arbitrary. However, just
as for the BTZ model, one can precisely compensate for this discrepancy  
with the
Jacobian prescription of   Section I. Also of interest,  only for our
formalism does the logarithmic correction 
turn out to be independent of $d$ (in the limit of small box size);
 including the $d=3$ BTZ model.

{\bf (ii) $L\sim R$}: 
In this particular regime, an immediate issue is the 
priorly discussed {\it stability} condition; namely, $\EPP>0$. 
From Eq.(\ref{144}), we find the following form for
the relevant quantity:
\be
E_0^{\prime\prime} = 
 {\LL^{d-3}\over (d-2)^2\omega_d}
\left[(d-1){\LL^2\over L^2}n_0^{-{d-3\over d-2}}
-(d-3) n_0^{-{d-1\over d-2}}\right]\;.
\label{151}
\ee
Evidently, stability directly implies a maximal  value for
$L$:
\be
L<L_{max}\equiv \sqrt{d-1\over d-3}R =
\sqrt{d-1\over d-3}\LL n_0^{1\over d-2}\;.
\label{152}
\ee
Not coincidentally,  $L=L_{max}$ can be identified as
the $d$-dimensional analogue of the Hawking-Page 
phase-transition point \cite{haw4}.

It is interesting to determine the entropic correction
as $L_{max}$ is approached by $L$ from below.
In this regard, it is, actually, more appropriate to
keep $L$ as the fixed parameter and let $n_0$ approach
its minimal value from above. That is, we can first
translate Eq.(\ref{152}) into
\be
n_0 > n_{min}\equiv\left[\sqrt{d-3\over d-1} {L\over\LL}\right]^{d-2}\;,
\label{blah} 
\ee
and then, in the spirit of a perturbative expansion, write
\be
n_0=n_{min}+\delta_{n}\;,
\label{153}
\ee
where $\delta_n$ is a small ({\it i.e.}, $\delta_{n}<< n_{min}\sim n_0$) 
but strictly 
positive  integer. 
Some straightforward calculation
then yields  (to lowest order in $\delta_n$):
\be
\EPP\sim \delta_n n_{0}^{-{2d-3\over d-2}} \quad {\rm and} \quad 
\beta^{-1}\sim \EP\sim n_{0}^{-{1\over d-2}}\;.
\label{155}
\ee
This means that
$\beta\EPP \sim \delta_n n_{0}^{-2}\sim S_{BH}^{-2}$ and
 Eq.(\ref{2}) takes on the form 
\be
S_C\approx S_{BH}+\ln[S_{BH}]\;.
\label{157}
\ee
Although this is a different result than found
in the prior (small box) limit, it is noteworthy
that the prefactor  still does {\it not} depend  
on the dimensionality of the spacetime.
In fact, this  ``blissful''  ignorance  of $d$
turns out to be a resilient feature of both the
canonical and grand canonical frameworks
(at least for the special cases considered
in this paper).  

A further comment regarding stability, near the
phase-transition point,
 is in order.
The logarithmic prefactor   of +1 is rather large
and implies that the thermal fluctuations 
are similarly large. (This can  be directly verified 
via Eq.(\ref{6}): $\Delta n\sim n_0$.)
Hence, when $\delta_n\sim {\cal O}(1)$, there
will be no means of suppressing  a phase transition and
the system is, actually,   only in  a {\it meta-stable} state.
To achieve ``true'' stability, it 
is necessary to move sufficiently
far  from the phase-transition point so
that the ratio $\Delta n /n_0$  is 
substantially smaller than unity.
Without claiming to be rigorous, let us
suppose that the system is stable
as long as $\Delta n\sim n_0^{1/ 2}$.
In this case, it is not difficult
to show that the {\it effective}
transition point occurs close to
$\delta_n\sim n_0$, which is,
essentially,  a regime of small box size.

\subsection{AdS-Reissner-Nordstrom}

The prior canonical formalism can also be 
applied to the case of a charged black hole,
with the understanding that the black
hole is  immersed in a  heat
bath which contains no {\it free} charges. 
That is to say, the  black hole charge
can  legitimately be regarded as a {\it fixed}
parameter if (and only if) there is no possibility for
the emission or absorption  of charged particles ({\it e.g.}, if the
temperature is smaller than the bare mass of an electron).
The  more interesting case of a  black
hole with a fluctuating charge will be the subject
of the following two sections.

When the charge, $Q$, is non-vanishing, the previous 
AdS horizon relation (\ref{139})
takes on a more  general  form \cite{mys},
\be
f(R;Q)\equiv {R^2\over L^2} +1-{\omega_d E\over R^{d-3}}
+{\omega_d^2 Q^2\over R^{2(d-3)}}
=0\;.
\label{158}
\ee
Solving for the energy, we now have
\be
E(R;Q)={1\over \omega_d}\left[{R^{d-1}\over L^2}+R^{d-3}
+{\omega_d^2Q^2\over R^{d-3}}\right]\;.
\label{159}
\ee

Since the relation between the entropy and $R$ 
is the same as before,
we can directly apply Eq.(\ref{142}) to obtain
the desired spectral form,
\be
E(n)= {1\over \omega_d}\left[{\LL^{d-1}\over L^2}n^{{d-1\over d-2}}
+\LL^{d-3} n^{{d-3\over d-2}} +{\omega_d^2 Q^2\over\LL^{d-3}}
n^{-{d-3\over d-2}}
\right]\;.
\label{160}
\ee
As in the prior subsection, we will concentrate
on certain limiting cases for which the analysis
somewhat simplifies. 

{\bf (i) $R>> L$  and $Q\sim 0$}:
This limit is essentially   case (i) of the prior subsection
and the calculations  need not be repeated.

{\bf (ii) $R >> L$  and  $Q\sim Q_{ext}$}:
Here, we have used $Q_{ext}$ to denote the charge of an
extremal black hole. It is of significance that the extremal  limit 
coincides with 
the limit of vanishing  temperature  ({\it i.e.}, $\beta^{-1}=0$).
Hence,  with  the natural assumption of
cosmic censorship, $|Q_{ext}|$  must  
also represent an upper bound on the magnitude of the 
charge.\footnote{It remains a point of controversy, in the literature, as to
how an extremal black hole should be interpreted thermodynamically
(see \cite{kun2x} for discussion and references).
Since our analysis formally  breaks down at
$\beta^{-1}=0$, we will be unable  to
address this particular issue.}

When delving into any regime of substantial charge,
we must necessarily consider the following pair
of stability constraints:
\be
 \beta^{-1}={\EP\over \epsilon}>0 \quad {\rm and}    \quad
\EPP>0\;.
\label{161}
\ee
Hence, let us be more precise with regard to the
quantities in question (keeping in mind that the $R>>L$ limit
is in effect):
\be
\EP \approx 
{1\over (d-2)\omega_d}\left[(d-1){\LL^{d-1}\over L^2}n_0^{{1\over d-2}}
-(d-3){\omega_d^2 Q^2\over\LL^{d-3}}
n_0^{-{2d-5\over d-2}}
\right]\;,
\label{162}
\ee
\be
\EPP \approx {1\over (d-2)^2\omega_d}\left[(d-1){\LL^{d-1}\over L^2}
n_0^{{-{d-3\over d-2}}}
+(d-3)(2d-5){\omega_d^2 Q^2\over\LL^{d-3}}
n_0^{-{3d-7\over d-2}}
\right]\;.
\label{163}
\ee

Since $\EPP$ is manifestly positive, we can  focus our
attention on just the condition $\EP>0$. This inequality implies
 a maximal value of $|Q|$; namely, the extremal
value of the charge. More specifically, the following constraint
must be imposed:
\be
Q^2<Q^2_{ext}\equiv\left({d-1\over d-3}\right){R^{2(d-2)}\over L^2 \omega_d^2} 
= \left({d-1\over d-3}\right){\LL^{2(d-2)}\over L^2 \omega_d^2} n_0^2\;.
\label{164}
\ee

Following the methodology of the prior subsection
({\it cf}, case (ii)),  let us
rather view Eq.(\ref{164}) as a lower bound on $n_0$ 
and then adopt the  perturbative form 
\be
n_0= n_{min} +\delta_n\;,
\label{165}
\ee
where $n_{min}$ is, now, the
lower bound on $n_0$ as dictated by Eq.(\ref{164})
and $\delta_n$ is, again, a small but strictly positive integer.

Substituting this relation into Eqs.(\ref{162},\ref{163}),
we obtain (to lowest order in $\delta_n$)
$\EPP\sim  n_0^{-{d-3\over d-2}}$,  $\beta^{-1}\sim\EP\sim
\delta_n n_0^{-{d-3\over d-2}}$, and so 
$\beta\EPP\sim {\rm constant}$.
Since constant terms in any entropy expression
can be safely discarded, the  canonical entropy
(\ref{2}) can now be written  as
\be
S_C\approx S_{BH}+{\cal O}[S_{BH}^{-1}]\;.
\label{169}
\ee
That is, the logarithmic correction has
been completely suppressed for a near-extremal black hole (in a small box).
In order to understand this phenomenon, let us also consider the  fluctuations
in the area for this near-extremal regime.
According to Eq.(\ref{6}), $\Delta A\sim\sqrt{\EP /\EPP}\sim$ constant,
so that the area fluctuations are similarly suppressed. 
Which is to say, the same basic mechanism
(that suppresses the logarithmic correction)  provides a natural means for 
the enforcement of  cosmic censorship. 

{\bf (iii) $R<< L$ and } any viable {\bf $Q$}:
With an eye toward the stability constraints
of  Eq.(\ref{161}), let us
re-evaluate the pertinent derivatives for
the $R<<L$ limit:
\be
\EP \approx {(d-3)\over (d-2)\omega_d}\left[\LL^{d-3}n_0^{-{1\over d-2}}
-{\omega_d^2 Q^2\over\LL^{d-3}}
n_0^{-{2d-5\over d-2}}
\right]\;,
\label{170}
\ee
\be
\EPP \approx {(d-3)\over (d-2)^2\omega_d}\left[
(2d-5){\omega_d^2 Q^2\over\LL^{d-3}}
n_0^{-{3d-7\over d-2}}
-\LL^{d-3}n_0^{-{d-1\over d-2}}
\right]\;.
\label{171}
\ee

Neither of the above  quantities is manifestly
positive, leading to  both an
upper   and a lower bound on the magnitude
of the charge.\footnote{Notably, this
point of minimal  charge can be identified
with   a Reissner-Nordstrom phase transition
that was first discussed by Davies \cite{DAV}.}  Quantitatively, these  are
given by
\be
Q^2 < Q^2_{ext}\equiv {R^{2(d-3)}\over \omega^2_d}
= {\LL^{2(d-3)}\over \omega^2_d} n_0^{2(d-3)\over d-2}\;,
\label{172}
\ee
\be
Q^2 > Q^2_{min}\equiv {R^{2(d-3)}\over (2d-5)\omega^2_d}
= {\LL^{2(d-3)}\over (2d-5) \omega^2_d} n_0^{2(d-3)\over d-2}\;.
\label{173}
\ee
Apparently, there is only a  small
range of charge values for which the black hole
(in a large box) can be stable. This is, however,  not much
of a surprise, given that a neutral black
hole has {\it no} chance for stability when
$L>> R$. 

Similarly to some previous cases,
we can  perturbatively expand $n_0$ about its minimal (maximal)
value and, thereby,   
 determine the logarithmic correction
as the  near-extremal (near-minimal) value of charge 
is {\it effectively} approached.  Here, we will simply quote
the final results:  
\be 
S_C\approx S_{BH} +{\cal O}\left[S_{BH}^{-1}\right]\;,
\label{184}
\ee
\be 
S_C\approx S_{BH} +\ln[S_{BH}]\;,
\label{185}
\ee
for the cases of near-extremal and  near-minimal
charge respectively.
Once again, we see that the logarithmic correction 
has been completely suppressed for a near-extremal
black hole. As previously discussed, this  effect  can be viewed
as a natural means of enforcing cosmic censorship.
On the other hand, the logarithmic correction (and, therefore, the
magnitude of the 
 fluctuations) is rather large  when the charge
is close to its  near-minimal value.
Since the allowed range of $|Q|$ is actually quite small,
 these large fluctuations have  severe implications
for the stability of the system (see the discussion
at the very end of the prior subsection).
It would seem that, for a black hole
in a large box, any state of thermal equilibrium  would
be, at best, a precarious situation of  meta-stability.
This point of view will be put on firmer ground
when we revisit the large-box scenario
in Section V.

\section{Grand Canonical Ensemble: General}

In this section, we will  rederive  
the canonical formalism of Section II under
the premise of a black hole
with a fluctuating charge. That is, it will now be assumed
that the thermal bath   
 contains charged
particles which can freely  interact with the black hole.
Although  the  charge ($Q$) can no
longer be regarded as a fixed quantity,  the
net charge of the black hole - that is, the
ensemble average of $Q$ - can still be zero.
Indeed, it is this case of a net vanishing
(but still fluctuating) charge that is the most
interesting from a physically motivated perspective. 
 
Given that there are now two fluctuating, independent
spectral parameters, it is most appropriate
(if not essential) to 
 model 
the system as a grand canonical ensemble.
In a conventional textbook sense, one can view the
charge  as a particle number, with some
suitable chemical (or, actually, electric) potential, $\mu$, 
relating the charge with the other thermodynamic parameters.  

With the above discussion in mind, we propose
that the partition function (\ref{119}) should now
be revised in the following manner:
\be
{\cal Z}_G(\beta,\mu)=\sum_{m=-\infty}^{\infty}\sum_{n=0}^{\infty} 
g_{n,m} \exp\left(-\beta \left[E(n,m)-\mu Q(m)\right]\right)\;.
\label{9}
\ee
Most importantly, we have introduced a   ``new'' quantum number, $m$, 
which directly measures
the black hole charge in accordance with
\be
Q=me\;,
\label{186}
\ee
where $e$ is some fundamental unit of electrostatic charge.
For future convenience, let us also define
\be
\lambda\equiv \beta\mu e\;.
\label{187}
\ee

To proceed,
it is first necessary to  specify some form for the degeneracy, $g_{n,m}$.
We will continue, for definiteness,  to  assume an uniformly 
spaced area spectrum. (For motivation in the case of
charged black holes, see \cite{kun2,kun3,das1,gj2,mak1}.) 
The black hole area law  can then be utilized
to fix this degeneracy up to a proportionality
constant.

Given that the area spectrum is evenly spaced, it can consequently
be deduced that
$A(n,m)\sim n+\alpha m^p$ ($n$,$|m|=0,1,2,...$),
 where $p$ is a rational (positive) number and $\alpha$
is a positive constant.
(That this spectral form is the correct generalization of $A\sim n$
 can  intuitively be seen from an 
 inspection of Eqs.(\ref{164},\ref{172}); also see
\cite{kun2,kun3,das1,gj1,gj2}. 
 These equations demonstrate that  $Q_{ext}\sim A$ for $R\gg L$ and 
$Q_{ext}\sim A^{(d-3)/(d-2)}$ for $R\ll L$. Therefore, since  
the quantum numbers,  $n$ and $m$, are supposed to be  independent,
one must necessarily let $n\rightarrow 0$ in the extremal limit
and then fix the $m$ dependence - that is, fix the power $p$ - accordingly. 
For instance,
Eqs.(\ref{164},\ref{172}) immediately imply that  $p=1$ when
$R\gg L$ and $p=\frac{d-2}{d-3}$ when $R\ll L$.)
Employing
the usual statistical interpretation of  entropy, 
$ g_{n,m} \propto e^{S(n,m)}$, as well as the area law, we then have
\be 
\ln g_{n,m}=\epsilon (n+\alpha m^p)\;,
\label{189}
\ee
where $\epsilon$ is, as before, a dimensionless
(positive) parameter of the order unity. 

Putting everything together and taking the continuum limit,
we can re-express
the grand canonical partition function (\ref{9})
as follows:
\be
{\cal Z}_G(\beta,\lambda)=\int_{-\infty}^{\infty}dm\int_{0}^{\infty}
dn 
\exp\left(-\beta E(n,m) +an+bm^p +\lambda m\right)\;,
\label{192}
\ee
where $a\equiv\epsilon$ and  $b\equiv\alpha\epsilon$. 

For calculational convenience, let us introduce
the following spectral function:  
\be
G(n,m)\equiv \beta E(n,m)-\XXXX\;,
\label{193}
\ee
which can also be expressed as an expansion  about the ensemble
averages ($n_0\equiv <n>$ and $m_0\equiv <m>$): 
\bea
G(n,m)=\GG + \left(n-n_0\right)\GP +\left(m-m_0\right)\GD
+{1\over 2}\left[\left(n-n_0\right)^2\GPP \right. 
\nonumber \\
\left. +\left(m-m_0\right)^2 \GDD +2\left(n-n_0\right)\left(m-m_0\right)
\GDP\right] +... \;.
\label{194}
\eea
Here (and for the duration), a prime/dot indicates a derivative with
respect to $n$/$m$, whereas a subscript of 0 represents a quantity
evaluated at $n=n_0$ and $m=m_0$ ({\it i.e.}, at thermal equilibrium).

Rewriting the exponent of Eq.(\ref{192}) in terms of
$G$ and   shifting  the integration variables
($x=n-n_0$, $y=m-m_0$), we have
\begin{eqnarray}
& & \ZG(\beta,\lambda)\approx 
\exp\left(-\GG +\lambda m_0\right)\times\nonumber\\
& & \int^{\infty}_{-\infty}dy\int^{\infty}_{-n_0}dx
\exp\left(-\left[\GP x + \GD y +{1\over 2}\GPP x^2
+{1\over 2} \GDD y^2 + \GDP xy-\lambda y\right]\right)\;.
\label{197}
\end{eqnarray}

Let us first consider the integration with respect to $x$.
Applying another coordinate shift and
taking  the semi-classical limit,  we obtain a Gaussian form,
\begin{eqnarray}
& & \ZG(\beta,\lambda)\approx \exp\left(-\GG +\lambda m_0\right)
\times\nonumber\\
& & \int^{\infty}_{-\infty}dy\int^{\infty}_{-\infty}dz
\exp\left(-{\GPP\over 2}z^2 
+{1\over 2 \GPP}\left[\GP+\GDP y\right]^2 -\left[{1\over 2}\GDD y
+\GD -\lambda\right]y\right)\;,
\label{198}
\end{eqnarray}
which can be readily integrated to give
\be
\ZG(\beta,\lambda)
\approx \exp\left(-\GG +\lambda m_0 +{(\GP)^2\over 2\GPP}\right)
\sqrt{2\pi\over \GPP}
\int^{\infty}_{-\infty}dy
\exp\left(-{1\over 2}\left[\Theta y^2 + 2\Phi y\right]\right)\;,
\label{199}
\ee
where
\be
\Theta\equiv \GDD-{(\GDP)^2\over \GPP}\;,
\label{200}
\ee
\be
\Phi\equiv \GD-{\GP\GDP\over \GPP}-\lambda\;.
\label{201}
\ee

The surviving integrand can also be rearranged (after
a coordinate shift) to reveal
another Gaussian, 
\be
\ZG(\beta,\lambda)\approx \exp\left(-\GG +\lambda m_0 +{(\GP)^2\over 2\GPP}
+{1\over 2}{\Phi^2\over \Theta}\right)
\sqrt{2\pi\over \GPP}
\int^{\infty}_{-\infty}dw
\exp\left(-{\Theta\over 2}w^2\right)\;,
\label{202}
\ee
and so we finally obtain
\be
\ZG(\beta,\lambda)\approx \exp\left(-\GG +\lambda m_0 +{(\GP)^2\over 2\GPP}
+{1\over 2}{\Phi^2\over \Theta}\right)
{2\pi\over\sqrt{ \GPP\GDD-(\GDP)^2}}\;.
\label{203}
\ee

The grand canonical entropy, $S_G$, can be evaluated with
the obvious analogue of Eq.(\ref{128}):
\be
S_{G}=\beta\langle E\rangle -\lambda\langle m\rangle+\ln\ZG
\approx \YYYY+\beta\left(\langle E\rangle -\E\right)-{1\over 2}\ln
\left[\GPP\GDD-(\GDP)^2\right]\;,
\label{ent}
\ee
where we have applied the explicit  forms of
$G$ (\ref{193}) and $\ZG$ (\ref{203}).
In realizing this expression, we  have also incorporated
the following equilibrium conditions:
\be
\GP=0\;,
\label{11}
\ee
\be
\GD=\lambda
\label{12}
\ee
(and, therefore, $\Phi=0$).
One can deduce these constraints
by first identifying  
the microcanonical free energy ({\it cf}, 
Eqs.(\ref{192},\ref{193})), $F(n,m)=G(n,m)-\lambda m$,
and then setting $F^{\prime}_0={\dot F}_0=0$.

For the purpose of simplifying the above result for $S_G$, it is
useful to consider the  ensemble average of the energy, 
\be
\langle E\rangle =-\frac{\partial}{\partial\beta}\ln\ZG
\approx \E+{1\over\beta}+{p(p-1)bm_{0}^{p-2}\GPP\over 
2\beta(\GPP\GDD-(\GDP)^2)}
\ee
(where we have applied the equilibrium conditions~(\ref{11},\ref{12}),
but only after differentiating with respect to $\beta$),
as well as
\be
\langle m^p\rangle =\frac{\partial}{\partial b}\ln\ZG
\approx m_{0}^{p}+{p(p-1)m_{0}^{p-2}\GPP\over 2(\GPP\GDD-(\GDP)^2)}\;.
\ee
It can then be shown that
 Eq.(\ref{ent}) reduces to
\be
S_{G}= S_{BH}-{1\over 2}\ln
\left[\GPP\GDD-(\GDP)^2\right]+{\cal O}(S_{BH}^{-1})\;,
\label{gent}
\ee
where we have identified
\be
S_{BH}={1\over 4G}\langle A(n,m) \rangle =an_{0}+b\langle m^{p}\rangle\;.
\ee

The logarithmic term in~(\ref{gent}) can now be recognized
as the leading-order thermal correction to the  black hole
entropy.
A more explicit form for the
grand canonical entropy (\ref{gent}) is the following:
\be
S_G \approx S_{BH} 
-{1\over 2}\ln
\left[{\EPP\left(\EDD-\alpha p(p-1)m_{0}^{p-2} \EP\right)
-(\EDP)^2\over (\EP)^2}
\right]\;.
\label{13}
\ee
Along with Eq.(\ref{193}) for $G(n,m)$, we have also
applied
\be
\beta={a\over \EP}={\epsilon\over \EP}\;,
\label{14}
\ee
\be
\lambda=\beta\ED-bp m_{0}^{p-1}=\beta\ED-\epsilon\alpha p m_{0}^{p-1}\;,
\label{15}
\ee
with these relations following directly from
Eqs.(\ref{11},\ref{12}).

%

As in the prior canonical treatment,
there are issues of stability  that need to 
 be addressed. For the current analysis,
 the procedure suffers a formal breakdown
({\it i.e.}, thermal equilibrium  can not be realized)
when  one or both of the following conditions
is violated:
\be
\EPP\left(\EDD-\alpha p(p-1)m_{0}^{p-2}\EP\right)
-(\EDP)^2 >0\;,
\label{17}
\ee
\be
\EP>0\;.
\label{18}
\ee
The first constraint is necessitated by the positivity 
of the  logarithmic argument, whereas the second follows
from  the positivity of the temperature
({\it cf}, Eq.(\ref{14})).

Before proceeding to the next section, we 
will consider the  thermal fluctuations
in  the spectral numbers, $n$ and $m$.  
As a consistency check,
let us first  calculate the thermal expectation values 
of these quantum numbers.
For this purpose, we can call upon the following relations:
\be
<n>= {\partial\ln \ZG \over \partial a}\;,
\label{222}
\ee
\be
<m>= {\partial\ln \ZG \over \partial \lambda}\;.
\label{228}
\ee

Let us, for the time being, concentrate on the quantum
number $n$.
Substituting Eq.(\ref{203}) for $\ZG$, we  
find
\be
<n> =n_0 +{\GD\GDP-\GP\GDD-\lambda \GDP\over \GPP\GDD-(\GDP)^2}\;.
\label{224}
\ee
To obtain this result, it  should be kept in mind that
only $\GG$ and $\GP$ depend on $a$ 
(such that $\partial_a \GG = -n_0$, $\partial_a \GP =-1$). 
It is now a simple matter to confirm that $<n>=<n_0>$
by virtue of the equilibrium conditions (\ref{11},\ref{12}).

In direct analogy to Eq.(\ref{5}), it is clear that
\be
(\Delta n)^2={\partial^2\ln\ZG\over\partial a^2}\;,
\label{226}
\ee
and this calculation yields
\be
(\Delta n)^2 ={\GDD\over \GPP\GDD-(\GDP)^2}\;,
\label{227}
\ee
or, equivalently,
\be
(\Delta n)^2 = {\EP\over \epsilon}
\left[{\EDD-\alpha p(p-1)m_{0}^{p-2}\EP\over
\EPP\left(\EDD-\alpha p(p-1)m_{0}^{p-2}\EP\right)-(\EDP)^2}
\right]\;.
\label{19}
\ee

The same general procedure reveals that $<m>=m_0$ and
\bea
(\Delta m)^2 = {\partial^2\ln\ZG\over\partial \lambda^2}
&=&{\GPP\over \GPP\GDD-(\GDP)^2}
\nonumber \\
&=& {\epsilon^{-1}\EP\EPP
\over\EPP\left(\EDD-\alpha p(p-1)m_{0}^{p-2}\EP\right)-(\EDP)^2}\;.
\label{232}
\eea


\section{Grand Canonical: Examples}

In this section, we will give an explicit
demonstration of the grand canonical formalism
by revisiting the AdS-Reissner-Nordstrom 
black hole of Section III(C). Let us re-emphasize, that
the prior (fixed charge) approach
is valid for a black hole immersed in a neutral
heat bath; otherwise, under more general circumstances,  
the current (fluctuating charge)
methodology is the appropriate one.

Let us begin by comparing the spectral form
of the entropy ({\it cf}, Eq.(\ref{189})),
$S(n,m)=\epsilon\left(n+\alpha m^2\right)$, with
the black hole area law, 
Eq.(\ref{141}).\footnote{As always, we are disregarding possible corrections 
that might appear in a complete theory of quantum gravity. To reiterate,
our current  interest  is in calculating {\it only} those deviations 
(from the Bekenstein-Hawking 
entropy)  that arise due  to thermal fluctuations.} 
In this way, we can express the horizon radius
as
\be
R^{d-2}(n,m)= \LL^{d-2} \left(n+\alpha m^p\right)\;.
\label{237}
\ee

Substituting the above result (and $Q=me$) into Eq.(\ref{159}) for
$E(R,Q)$,
we obtain the updated spectral form of the energy,
\begin{eqnarray}
E(n,m) & = & 
\frac{1}{\omega_d}\left[\frac{\LL^{d-1}}{L^2}\BBB^{\frac{d-1}{d-2}}
+\LL^{d-3} \BBB^{\frac{d-3}{d-2}} +\frac{\omega_d^2 e^2 m^2}{\LL^{d-3}}
\BBB^{-\frac{d-3}{d-2}}
\right]\nonumber\\
&\equiv& f(\A)+m^2 g(\A)\;,
\label{239}
\end{eqnarray}
where $\A\equiv \BBB$ is 
the dimensionless area. 

In order to simplify the upcoming analysis,
 we will also make use of the following notation:
\be
\QQ\equiv\EPP\left(\EDD-\alpha p(p-1)m_{0}^{p-2}\EP\right)-(\EDP)^2
\ee
or,  in terms of $f(\A)$ and $g(\A)$,
\be
\QQ =2g_0f_0^{\prime\prime}
+2m_0^{2}g_0g_0^{\prime\prime}-4m_0^{2}(g_0^{\prime})^{2}\;.
\label{ddd}
\ee

We are now well positioned for some explicit calculations.
As in Section III, the focus will be  on certain limiting
cases for which  the analysis is  most tractable.

{\bf (i) $R>> L$ and $Q\sim 0$}: 
First note that, in the small $L$ limit,  we must choose  $p=1$ 
if the quantum numbers, $n$ and $m$, are to be independent (see the 
discussion leading up to Eq.(\ref{189})). 
 However, as  shown below,
it is, for the purpose of deducing the logarithmic correction,
actually {\it not} necessary 
that $p$ be explicitly fixed.

In this  case of small (effective) box size,  the 
middle term in Eq.(\ref{239}) can be disregarded. Furthermore,
we will eventually  take $m_0\rightarrow 0$, but only at the end
of  each calculation. (Even if $m_0=0$, the quantum number
$m$ is free to fluctuate.)
Some  useful expressions include
\be
f_0^{\prime} \approx {(d-1)\LL^{d-1}\over (d-2)\omega_d L^2}
\A_0^{1\over d-2}\;,\;\;\;
g_0^{\prime}\approx
-\frac{(d-3)\omega_d e^2}{(d-2)\LL^{d-3}} m_0^2\A_0^{-{2d-5\over d-2}}\;,
\label{240}
\ee
\be
f_0^{\prime\prime}  \approx {(d-1)\LL^{d-1}\over (d-2)^2\omega_{d} L^2}
\A_0^{-{d-3\over d-2}}\;,\;\;\;
g_0^{\prime\prime}\approx\frac{(d-3)(2d-5)\omega_d e^2}{(d-2)^{2} \LL^{d-3}}
m_0^2\A_0^{-{3d-7\over d-2}}\;,
\label{241}
\ee
\be
\QQ\approx 
2 {e^2\over (d-2)^2}\left[
(d-1){\LL^2\over L^2}\A_0^{-2{d-3\over d-2}}
+(d-3) {\omega^2_de^2\over \LL^{2(d-3)}} m_0^2\A_0^{-2{{2d-5}\over d-2}}
\right]\;.
\label{242}
\ee


First, let us calculate the logarithmic correction to the
entropy; {\it cf}, Eq.(\ref{13}). In the 
limit of vanishing $m_0$, we obtain 
\be
\ln \left[{\QQ\over (\EP)^2}\right]\approx
-2\ln\A_0\;.
\label{244}
\ee
Hence, the grand canonical entropy (\ref{13}) can be written as
\be
S_G\approx S_{BH}+\ln[S_{BH}]\;.
\label{255}
\ee

That the logarithmic prefactor is now equal to
+1 is quite an intriguing outcome. Recall that, 
for a black hole with a fixed charge (in the exact same limit), we found
a value of +1/2 ({\it cf}, Eq.(\ref{150})). This implies that
each  quantum number ({\it i.e.}, each freely
fluctuating parameter) induces a thermal correction to  the entropy
of precisely ${1\over 2}\ln S_{BH}$.  It would have been difficult
to advocate such an outcome beforehand,
inasmuch as  $n$ and $m$ make (in general) inequivalent contributions  to the 
area spectrum. It is also worth noting that Major and Setter \cite{ms2x}
found the same prefactor of +1 in their variant of the grand canonical
ensemble. This agreement (in related but distinct methods) 
further suggests that the value of +1/2  per quantum number
is a resilient result.

Next, let us calculate the quantum fluctuations in the spectral numbers,
$n$ and $m$.
Substituting the relevant formalism
into Eqs.(\ref{19},\ref{232}), we ultimately find that, for
the limiting case of current interest, 
\be
(\Delta n)^2 \sim (\Delta m)^2 \sim \A_0 \;.
\ee  
Given the presumed choice of $p=1$,
it follows that
$\Delta S_{BH}\sim \Delta n + \Delta m \sim S_{BH}^{1\over 2}$,  in
compliance with the intuitive expectations.\footnote{Note that
our finding of $(\Delta Q)^2\sim S_{BH}$ seems to be in 
conflict with the results
of a previous  study \cite{sch}, which found $(\Delta Q)^2\sim $ constant.
However, \cite{sch} assumes  a  stable, neutral black hole
in a  large-sized box;  a scenario 
that turns out to be disallowed by our formalism (see case {\it iii} 
below).}

{\bf (ii) $R>> L$ and $Q\sim Q_{ext}$ }:
Let us begin here by recalling that the choice
of $p=1$ is still the appropriate one.
Hence, the relation between the extremal values of $\A$ and $m$ 
is simply  $\A _{ext}=\alpha m_{ext}$.
Given this observation, we can now use   perturbative
techniques (following the general procedure outlined in
Section III)
to calculate the logarithmic correction
for a near-extremal black hole.
Utilizing the  relevant formalism (\ref{240}-\ref{242})
and appropriately expanding  $m_0$  (and $\A_0\sim \alpha m_0$) 
just below  $m_{ext}$, 
we find that,  near extremality,
\be
\EP=f_0^{\prime}+m_0^{2}g_0^{\prime\prime}
\sim \delta_m \A_0^{-{d-3\over d-2}}\;,
\label{adad}
\ee
\be
\QQ\sim\A_0 ^{-2\frac{d-3}{d-2}}\;,
\label{ada}
\ee
where $\delta_{m}$ is a small but strictly positive integer
that approaches zero in the extremal limit.

Now substituting
 Eq.(\ref{adad}) and Eq.(\ref{ada}) into Eq.(\ref{gent}), 
we obtain just as before (in the analogous case with a fixed charge),
\be
S_C\approx S_{BH} + {\cal O}[S_{BH}^{-1}]\;.
\label{269}
\ee

It can also be shown  that, near extremality,
 both the area and charge fluctuations are completely
suppressed; that is, $\Delta n\sim\Delta m\sim$ constant.  
To put it another way,
the fluctuations ``freeze'' as extremality is approached and
cosmic censorship  can {\it not}
be violated by a fluctuating geometry.


{\bf (iii) $R<< L$ and } any viable {\bf $Q$}:
As it turns out, there is no viable $Q$ in this limit of large
(effective) box size; that is, in 
the limit of an asymptotically flat spacetime.
To demonstrate this oddity, let us first recall  the two
stability conditions for a grand canonical ensemble (\ref{17},\ref{18}).  
Hence, it is appropriate to consider
the following quantities (in the $L>>R$ limit):
\be
\EP \approx {(d-3)\over (d-2)\omega_d}\left[\LL^{d-3}\A_0^{-{1\over d-2}}
-{\omega_d^2 e^2m_0^2\over\LL^{d-3}}
\A_0^{-{2d-5\over d-2}}
\right]\;,
\label{274}
\ee
\be
\QQ \approx
{2(d-3)e^2\over (d-2)^2 \LL^{2(d-3)}}\left[\omega_d^2 e^2 m_0^2
\A_0^{-2{2d-5\over d-2}} 
- \LL^{2(d-3)}
\A_0^{-2}
\right]\;.
\label{275}
\ee

Imposing positivity on  the above expressions,
we can directly extract the following pair of stability constraints:
\be
e^2 m^2_0<Q^2_{ext}={\LL^{2(d-3)}\over \omega_d^2}\A_0^{2{d-3\over d-2}}\;,
\label{276}
\ee
\be
e^2 m^2_0>Q^2_{min}={\LL^{2(d-3)}\over \omega_d^2}\A_0^{2{d-3\over d-2}}\;.
\label{277}
\ee
Obviously, it is impossible to simultaneously satisfy both of these conditions;
meaning that  stability can {\it never} be achieved 
(in the fluctuating-charge scenario)  when $L>>R$.
One caveat might be a {\it perfectly} extremal black hole, since such
an entity does not exchange heat with its surroundings
nor does it experience thermal fluctuations. 
It is, however, interesting to note
that these same properties will prohibit
an extremal black hole from continuously
evolving into a non-extremal black hole. Since the converse
process must, therefore, also be forbidden, we have
 an example of  the third law  of black hole mechanics
at work.

\section{Concluding Discussion}

In summary, we have been investigating  the effect of thermal
fluctuations on the  entropy of a  black hole.
The study focused on  the picture of a black hole
in a ``box''; with  the system  modeled both as
a canonical ensemble and a grand canonical ensemble,
depending on whether the black hole charge is fixed
or allowed to fluctuate (respectively). We were guided,
in large part, by the philosophy that the quantum
spectrum is an important ingredient 
in any  analysis that endeavors to consider the
corrections to the entropic area law. 
For definiteness, we chose to work, throughout, with
an uniformly spaced area spectrum, as this  spectral
form has considerable support in the literature.
It would be interesting, however, to see  
the repercussions on our results as the
spectrum deviates gradually from 
this (perhaps) idealized form.

Throughout the paper, the generic
formalism was punctuated with a number of specific  models. Hence, we
accumulated a wide array of interesting results;
both quantitative and qualitative. Let
us now summarize, in point-form, some of the
more prominent outcomes.\\
{\it (i)} The  leading-order correction to
the canonical or grand canonical entropy  can typically be expressed
as the logarithm of the classical entropy.  For many
interesting cases, the logarithmic prefactor is
a simple integer or half-integer that does {\it not}
depend on the dimensionality of the spacetime.
 \\
{\it (ii)} For an AdS-Schwarzschild black hole 
with $R>>L$, the logarithmic
prefactor was found to be +1/2, irrespective of the dimensionality.
(This includes the BTZ black hole, for any value of $L$.)
This value is notably in  conflict with
some of the pre-existing literature  
({\it e.g.}, \cite{maj3x}). \\
{\it (iii)} For an AdS-Reissner-Nordstrom black hole,
the calculations depend 
strongly on whether  the charge is
regarded as a fixed or fluctuating quantity. For instance,
if $R>>L$ and the charge is small, the  prefactor increases
from +1/2 to +1   when the fluctuations are ``turned on''.
This larger value does happen to 
agree, precisely,  with an earlier treatment on the grand canonical
ensemble \cite{ms2x}. \\
{\it (iv)} We have demonstrated that, for a black hole
which is far from extremality (and $R>>L$),
the quantum numbers labeling  the area spectrum
fluctuate (from their equilibrium values) 
according to $\Delta n$, $\Delta m  \sim S_{BH}^{1\over 2}$.
It can therefore be inferred that
 $\Delta A \sim \Delta Q \sim A^{1\over 2}$.
Note that, although $\Delta A$ and $\Delta Q$ are rather large, 
the relative  variations, 
$\Delta A/A\sim \Delta Q/Q_{ext}\sim A^{-{1\over 2}}$,
are quite small. \\
{\it (v)} When $L>>R$ ({\it i.e.}, the asymptotically flat-space
limit), the enforcement of stability severely
restricts the  solution space. For instance, if
the charge is a fixed quantity,  
stability seems unlikely and could {\it only} be possible
for   black holes that are very
close to extremality.
 Meanwhile, when the charge fluctuates,
 stability can {\it not}  possibly be achieved  for any non-extremal black
hole.   

As stressed in the early parts of the paper,
we have been  neglecting the  quantum corrections
to the entropy that arise at the microcanonical level.
Insofar as any current  theory
of quantum  gravity is, at best, a work in project,
there are  conceptual  limitations in 
attempts at   quantifying this microcanonical correction. 
Nonetheless,
certain calculations - especially, in the
context of loop quantum gravity  \cite{maj1x} -
suggest a microcanonical correction of $-{3\over 2}\ln S_{BH}$.   
If this value turns out be correct, then it  would
be perfectly valid  to simply subtract  off 3/2
from the  prefactor of the thermally induced
correction. (Higher-order corrections would, however,
be a substantially more complicated ordeal.)

A more serious omission in our formalism  was neglecting 
the fluctuations in spin. Unlike the case of
charge, in which the black hole can (in principle) be placed in an
electrostatic
heat bath, it is difficult to envision how the spin
fluctuations might possibly be suppressed. 
(It is not relevant as to whether   the black hole is, itself,
rotating  or stationary. The spin fluctuations should
still, at least naively, be of the order $S_{BH}^{1/2}$.
Indeed, some preliminary results \cite{gj3} have substantiated
that this estimate is correct.)
Hence, any eventual  discussion of physically realistic black holes
will have to  find a way of  incorporating these effects.
 Unfortunately, the vector
nature of any angular-momentum operator makes it highly
 non-trivial to extend our formalism in this direction.
Nonetheless, we can still make some rough estimates by
assuming that {\it (i)} the various spin components fluctuate
independently and {\it (ii)} each such component contributes
a quantum correction of   ${1\over2}\ln S_{BH}$ to the  grand canonical
entropy (see  the discussion following Eq.(\ref{255})).
Let us consider a (for instance)  four-dimensional black hole;
 this would imply a {\it maximal} thermal
correction of $5\times{1\over2}\ln S_{BH}$. Also subtracting off 
 the (estimated) microcanonical correction, we then have
 the following upper bound:
\be
S_{EI}\leq S_{BH}+ (1)\ln[S_{BH}]\;,
\label{555}
\ee
where the subscript $EI$ stands for ``everything included''. 
It is an intriguingly simple result, which we hope
to  readdress (much more rigorously) at a future time.

\section{Acknowledgments}

The authors graciously thank V.P. Frolov for
helpful conversations and J.D. Bekenstein for
his comments and suggestions. GG is also grateful for
the Killam Trust for its financial support.


\begin{thebibliography}{99}

\bibitem{smo} See, for a recent discussion, 
 L. Smolin, ``How far are we from the Quantum 
Theory of Gravity'' , hep-th/0303185 (2003).
\bibitem{bek1} J.D. Bekenstein, Lett. Nuovo. Cim. {\bf 4}, 737 (1972);
Phys. Rev. {\bf D7}, 2333 (1973); Phys. Rev. {\bf D9}, 3292 (1974).
\bibitem{haw1} S.W. Hawking, Comm. Math. Phys. {\bf 43}, 199 (1975).
\bibitem{wald1} See, for further discussion and references,
 R.M. Wald, Living Rev. Rel. {\bf 4}, 6 (2001)
[gr-qc/9912119].
\bibitem{str1}  A. Strominger and C. Vafa, 
Phys. Lett. {\bf B379}, 99 (1996) [hep-th/9601029].
\bibitem{ash}  A.Ashtekar, J. Baez, A. Corichi and
K. Krasnov, Phys. Rev. Lett. {\bf 80}, 904 (1998) [gr-qc/9710007].
\bibitem{maj1x} R.K.  Kaul and P. Majumdar, Phys. Rev. Lett. {\bf 84}, 5255
(2000) [gr-qc/0002040].
\bibitem{makx} J. Makela and P. Repo, ``How to interpret Black
Hole Entropy?'', gr-qc/9812075 (1998).
\bibitem{carx} S. Carlip, Class. Quant. Grav. {\bf 17}, 4175 (2000)
[gr-qc/0005017].
\bibitem{das2x} S. Das, P. Majumdar and R.K. Bhaduri, Class. Quant. Grav.
{\bf 19}, 2355 (2002)  [hep-th/0111001].
\bibitem{sol1x} S.N. Solodukhin, Phys. Rev. {\bf D51},
609 (1995) [hep-th/9407001]; {\it ibid}, 618 (1995) [hep-th/9408068];
{\it ibid} {\bf D57}, 2410 (1998) [hep-th/9701106].
\bibitem{furx} D.V. Fursaev, Phys. Rev. {\bf D51}, 5352 (1995) 
[hep-th/9412161].
\bibitem{loux} C.O. Lousto, Phys. Lett. {\bf B352}, 228 (1995)
[gr-qc/9411037].
\bibitem{frox} V.P. Frolov, W.Israel and S.N. Solodukhin,
Phys. Rev. {\bf D54}, 2732 (1996) [hep-th/9602105].
\bibitem{sol2x} R.B. Mann and S.N. Solodukhin,
Phys. Rev. {\bf D55}, 3622 (1997)
[hep-th/9609085];  Nucl. Phys. {\bf B523},
293 (1998) [hep-th/9709064].
\bibitem{kasx} H.A. Kastrup,  Phys. Lett. {\bf B413}, 267 (1997)
[gr-qc/9707009].
\bibitem{kun1x} A.J.M.  Medved and G. Kunstatter, Phys. Rev. {\bf D60},
104029 (1999) [hep-th/9904070].
\bibitem{gg1x} G. Gour, Phys. Rev. {\bf D61}, 021501 (2000)
[gr-qc/9907066].
\bibitem{ostx} O. Obregon, M. Sabido and V.I. Tkach, Gen. Rel. Grav.
{\bf 33},  913 (2001) [gr-qc/0003023].
\bibitem{jinx}
 J. Jing and M.-L. Yan, Phys. Rev. {\bf D63}, 024003 (2001) [gr-qc/0005105].
\bibitem{das1x}
 S. Das, R.K. Kaul and P. Majumdar, Phys. Rev. {\bf D63}, 044019 (2001)
[hep-th/0006211].
\bibitem{bir1x} 
D.Birmingham and S. Sen, Phys. Rev. {\bf D63}, 047501 (2001) 
[hep-th/0008051]. 
\bibitem{maj2x} P. Majumdar, Pramana {\bf 55}, 511 (2000)
[hep-th/0009008]; ``Black Hole Entropy: Certain Quantum
Features'', hep-th/0011284 (2000); ``Black Hole Entropy:
Classical and Quantum Aspects'', hep-th/0110198 (2001).
\bibitem{kun2x} A.J.M. Medved and G. Kunstatter, Phys. Rev. {\bf D63},
104005 (2001) [hep-th/0009050].
\bibitem{ms2x} S.A. Major and K.L. Setter, Class. Quant. Grav.
{\bf 18}, 5125 (2001) [gr-qc/0101031];
{\it ibid} 5293 (2001) [gr-qc/0108034].
\bibitem{bir2x} D. Birmingham, I. Sachs and S. Sen,
Int. J. Mod. Phys. {\bf D10}, 833 (2001) [hep-th/0102155].
\bibitem{govx}
T.R. Govindarajan, R.K. Kaul and V. Suneeta, Class. Quant. Grav. {\bf 18},
2877 (2001) [gr-qc/0104010].
\bibitem{cavx} M. Cavaglia and A. Fabbri, Phys. Rev. {\bf D65},
044012 (2002) [hep-th/0108050].
\bibitem{gup1x} K.S. Gupta and S. Sen, Phys. Lett. {\bf B526}, 121 (2002) 
[hep-th/0112041]. 
\bibitem{ajm1x} A.J.M. Medved, Class. Quant. Grav. {\bf 19}, 2503
(2002) [hep-th/0201079].
\bibitem{gup2x} K.S. Gupta, ``Near-Horizon Conformal
Structure and Entropy of Schwarzschild Black Holes'',
hep-th/0204137 (2002).
\bibitem{mukx} S. Mukherji and S.S. Pal, JHEP {\bf 0205}, 026
(2002) [hep-th/0205164].
\bibitem{das3x} S. Das, ``Leading Log Corrections to
Bekenstein-Hawking Entropy'', hep-th/0207072 (2002).
\bibitem{ajm2x} A.J.M. Medved,  ``Quantum-Corrected Entropy
for 1+1-Dimensional Gravity Revisited'', hep-th/0210017 (2002), and
to appear in Classical and Quantum Gravity;
 ``Horizon Dynamics of a BTZ Black Hole'', gr-qc/0211004 (2002).
\bibitem{gg2x} G. Gour, Phys. Rev. {\bf D66}, 104022 (2002)
 [gr-qc/0210024].
\bibitem{kripx} I.B. Khriplovich, ``How are Black Holes Quantized'',
gr-qc/0210108 (2002).
\bibitem{kaul1x} R.K. Kaul and S.K. Rama, ``Black Hole Entropy
from Spin One Punctures'', gr-qc/0301128 (2003).
\bibitem{kaul2x} R.K. Kaul ``Black Hole Entropy
from a Highly Excited Elementary String'', hep-th/0302170
(2003).
\bibitem{maj3x} A. Chatterjee and P. Majumdar, ``Black Hole
Entropy: Quantum versus Thermal Fluctuations'', gr-qc/0303030
(2003).
\bibitem{das4x} S. Das and V. Husain, ``Anti-de Sitter Black
Holes, Perfect Fluids and Holography'', hep-th/0303089
(2003).
\bibitem{das5x} M.M. Akbar and S. Das, ``Entropy Corrections
for Schwarzschild and Reissner-Nordstrom Black Holes'',
hep-th/0304076 (2003).
\bibitem{btz} M. Banados, C. Teitelboim and J. Zanelli, Phys. Rev. Lett.
{\bf 69}, 1849 (1992) [hep-th/9204099].
\bibitem{bek2} J.D. Bekenstein, Lett. Nuovo Cimento, {\bf 11},
467 (1974).
\bibitem{muk} V.F. Mukhanov, JETP Letters {\bf 44}, 63 (1986).
\bibitem{bek3} J.D. Bekenstein and V.F. Mukhanov, Phys. Lett. {\bf B360}, 7
(1995) [gr-qc/9505012];  J.D. Bekenstein, ``Quantum Black Holes
as  Atoms'',  in {\it Proceedings of the Eighth
Marcel Grossmann Meeting on General Relativity}, eds.
T. Piran and R. Ruffini (World Scientific, Singapore 1999),
pp. 92-111 [gr-qc/9710076]. 
\bibitem{bek5} J.D. Bekenstein, ``The Case for
Discrete Energy Levels of a Black Hole'',  in
{\it 2001: A Spacetime Odyssey}, eds. M.J. Duff and J.T. Liu
(World Scientific, Singapore 2002), pp. 21-31
 [hep-th/0107045].
\bibitem{bek6} J.D. Bekenstein and G. Gour, Phys. Rev.
{\bf D66}, 024005 (2002) [gr-qc/0202034].
\bibitem{kun1} A. Barvinsky and G. Kunstatter, ``Mass Spectrum
for Black Holes in Generic 2-D Dilaton Grvaity'', gr-qc/9607030
(1996).
\bibitem{kun2}  A. Barvinsky, S. Das and G. Kunstatter, 
Class. Quant. Grav. {\bf 18} 4845 (2001)
[gr-qc/0012066]. 
\bibitem{kun3} A. Barvinsky, S. Das and G. Kunstatter,
Phys. Lett. {\bf B517}, 415 (2001)
 [hep-th/0102061]; Found. Phys. {\bf 32} 1851 (2002)
[hep-th/0209039] (2002). 
\bibitem{das1} S. Das, P. Ramadevi and U.A. Yajnik,
Mod. Phys. Lett. {\bf A17}, 993 (2002) [hep-th/0202076];
S. Das, P. Ramadevi, U.A. Yajnik and A. Sule,
``Quantum Mechanical Spectra of Charged Black Holes'',
hep-th/0207169 (2002).
\bibitem{gj1} G. Gour and A.J.M. Medved, ``Kerr Black Hole as a
Quantum Rotator'', gr-qc/0211089 (2002), and to appear in
Classical and Quantum Gravity.
\bibitem{gj2} G. Gour and A.J.M. Medved, ``Quantum Spectrum for
a Kerr-Newman Black Hole'', gr-qc/0212021 (2002), and to appear
in Classical and Quantum Gravity.
\bibitem{mak1} J. Louko and J. Makela, Phys. Rev. {\bf D54},
4982 (1996) [gr-qc/9605058];
 J. Makela and P. Repo,
Phys. Rev. {\bf D57}, 4899 (1998) [gr-qc/9708029];
J. Makela, P. Repo, M. Luomajoki and J. Piilonen,
Phys. Rev. {\bf D64}, 024018 (2001) [gr-qc/0012005].
\bibitem{lgq} A. Alekseev, A.P. Polychronakos and M. Smedback,
``On the Area and Entropy of a Black Hole'', hep-th/0004036
(2000);  A.P. Polychronakos, ``Area Spectrum
and Quasinormal Modes of Black Holes'', hep-th/0304135 (2003).
\bibitem{wald2} See, for instance,
 R.M. Wald, {\it General Relativity} (University of
Chicago Press, 1984).
\bibitem{ads} See, for a review, A. Aharony, S.S. Gubser, J. Maldacena,
H. Ooguri and Y. Oz, Phys. Rept. {\bf 323}, 183
2000 [hep-th/9905111].
\bibitem{mys} See, for instance, 
 A. Chamblin, R. Emparan, C.V. Johnson and R.C. Myers,
Phys. Rev. {\bf D60}, 064018 (1999) [hep-th/9902170]. 
\bibitem{haw4} S.W. Hawking and D.N. Page, Commun. Math. Phys.
{\bf 87}, 577 (1983).
\bibitem{DAV} P.C.W. Davies, Proc. Roy. Soc. Lond. {\bf A353},
499 (1977).
\bibitem{sch} M. Schiffer, ``The Random Walks
of a Schwarzschild Black Hole'', gr-qc/9706060
(1997).
\bibitem{gj3} G. Gour and A.J.M. Medved, work in progress.

\end{thebibliography}
\end{document}